# Air-Stable Room-Temperature Quasi-2D Tin Iodide Perovskite Microlasers

Sangyeon Cho[1,2†], Wenhao Shao[3,4†], Jeong Hui Kim[3], Letian Dou[3,5,6*], Seok-Hyun Yun[1,2,7*]

[1]Wellman Center for Photomedicine, Massachusetts General Hospital, 65 Landsdowne St., Cambridge, MA 02139, USA

[2]Harvard Medical School, Boston, MA 02139, USA

[3]Davidson School of Chemical Engineering, Purdue University, West Lafayette, IN, 47907 USA

[4]Department of Chemistry, University of Georgia, Athens, GA 30602 USA

[5]Department of Chemistry, Purdue University, West Lafayette, IN, 47907 USA

[6]Department of Chemistry, Emory University; Atlanta, GA 30322, USA.

[7]Harvard-MIT Health Sciences and Technology, Cambridge, MA 02139, USA

[†]Equal Contribution

*Correspondence to: dou10@purdue.edu or syun@hms.harvard.edu

## Abstract

Quasi-2D tin iodide perovskites (TIPs) are promising lead-free alternatives for optoelectronic applications, but achieving stable lasing remains challenging due to their limited environmental stability. Here, we report air-stable, room-temperature lasing from quasi-2D TIP microcrystals as small as 4 μm. Incorporation of the organic spacer 5IPA3 significantly enhanced the stability of these materials compared to previously reported TIPs. Lasing was observed from both dielectric (n=4) and plasmonic (n=3 and n=4) TIP microlasers. Under picosecond pumping, lasing was sustained for over $10^8$ pump pulses in ambient conditions. These results represent a significant step toward practical photonic applications of tin-based perovskites.

## 1. Introduction

Lead halide perovskites (LHP) have attracted significant attention due to their exceptional performance in a range of optoelectronic technologies, including photovoltaics[1], light-emitting diodes[2], and lasers[3]. However, concerns about lead toxicity[4] and environmental risks[5] have spurred interest in alternative compositions. Among these, tin iodide perovskites (TIPs) have been proposed[6] and are currently under active investigation as leading lead-free candidates for optoelectronic applications[7]. TIPs offer even higher carrier mobility[8] and lower non-radiative losses[9] than their lead counterparts. TIP-based solar cells have achieved power conversion efficiencies exceeding 14%[10] and operational stability beyond 1,300 hours in nitrogen-purged environments.[11] Moreover, animal toxicity assays have shown no clear evidence of acute toxicity or genetic-level risks associated with TIPs.[12] Despite these advantages, their primary limitations remains the poor material stability and fabrication reproducibility, largely due to the high susceptibility of Sn(II) to oxidation into Sn(IV)[13] and poor thermal instabilities.[14] Oxidation susceptibility and low defect formation energy facilitate vacancy formation and accelerate degradation in the presence of air, moisture, and light.[15–17]

Recently, Ruddlesden-Popper perovskites, which consist of two-dimensional (2D) or quasi-2D perovskite slabs separated by cationic spacers, have attracted growing interest.[18-19] The quantum-well architecture imparts distinctive physical and optical characteristics, influencing charge transport[20,21], exciton dynamics[19,22], and ferroelasticity[23–25]. The spacer layers not only provide structural versatility but also help suppress oxygen permeation and improve photostability compared to 3D bulk counterparts.[20,21] Ruddlesden-Popper TIPs, incorporating various spacers and layer thicknesses, have been actively explored for lasing and related applications. Optically pumped lasing has been demonstrated from PEA-incorporated 2D TIPs using a high-Q distributed feedback resonator at 77 K.[26] Lasing from TIP microflakes incorporating diverse cations such as PEA, 2T+ (bithiophenylethylammonium), and 3T+ (2-([2,2':5',2"-terthiophen]-5-yl)ethan-1-aminium), have exhibited lasing at 83 K in nitrogen.[27] Similarly, lasing has been observed at 88 K in argon from ~100 μm-long TIP nanowires synthesized using BrCA3 (2-(2-bromo-5-carboxyphenoxy)ethan-1-aminium).[28] However, achieving stable lasing under ambient conditions has remained an unresolved challenge for TIPs.

In this study, we demonstrate room-temperature lasing in air from quasi-2D TIPs incorporating 2-(3,5-dicarboxyphenoxy)ethan-1-aminium (5IPA3) as the organic spacer. The 5IPA3 molecule, bearing two carboxyl groups, forms robust and highly directional hydrogen bonds with the perovskite slabs. This hydrogen-bonding network not only reduces oxygen permeability but also increases the mechanical rigidity of the lattice[29], which is believed to enhance thermal stability under high optical pumping conditions.[26,30] The resulting material, $(5IPA3)_2(MA)_{n-1}Sn_nI_{3n+1}$ (where $n$ represents the number of perovskite layers between organic spacers), exhibited substantially improved ambient stability both in the dark and under photoexcitation, compared to well-known 2D PEA-TIPs. Notably, we achieved, for the first time to our knowledge, air-stable room-temperature lasing from quasi-2D TIP microcrystals ($n$ = 3 and 4) under ambient conditions. This achievement represents a milestone in the development of lead-free perovskites and advances their potential for photonic and optoelectronic applications.

## 2. Results

### 2D tin iodide perovskites with 5IPA3 spacer

Benzoic acid was chosen as the molecular backbone of 5IPA3, as inspired by the recently developed molecular templating methods[28] to permutate the anisotropy in organic sublattices using robust and directional H bonded network. Specifically, the benzene-1,3-dicarboxylic acid core was introduced to create a sophisticated organic network to reduce the permeability of air and moisture. Details of the spacer design and crystal growth methods will be described elsewhere. Phase-pure quasi-2D perovskites exhibited trends that higher n values yield higher optical gain.[27,31] Figure 1a depicts the crystal structure of $(5IPA3)_2(MA)_3Sn_4I_{13}$ for $n$ = 4. The thickness of the organic sublattice is ~17.6 Å and that of the inorganic perovskite layers is ~24 Å. The dielectric permittivity of 5IPA3-n4 is approximately 4.[32] Considering the sub-nanometer-thick organic lattices, it is expected to allow sub-picosecond out-of-plane exciton transport via tunneling[33] and out-of-plane exciton diffusion lengths of ∼100 nm.[34] 5IPA3-n4 microcrystals were obtained by breaking down bulk crystals via sonication (Fig. 1b)[28]. The resulting microplates had a median length of 13 μm and a mean width of 2 μm (Supplementary Figs. S1 and S2). The median thickness of microplates was 545 nm, which corresponds to a total of 131 quantum wells.

Figure 1c presents the diffusion-reflectance spectra of microcrystals with different $n$ from 1 to 4. The quantum-well band edge energy was estimated from the onset of the absorption edges, which are 1.44 eV for 5IPA3-n4 ($n$ = 4), 1.54 eV for 5IPA3-n3 ($n$ = 3), 1.63 eV for 5IPA3-n2 ($n$ = 2), and 1.91 eV for 5IPA3-n1 ($n$ = 1). Stronger exciton binding in lower $n$ phases increases both higher Auger recombination rate[35] and exciton–phonon coupling.[36] Additionally, in the electron-hole plasma lasing scenario[37–39], the smaller exciton Bohr radius at lower $n$ leads to a higher carrier density for transparency.[40] Figure 1d shows photoluminescence (PL) spectra from microcrystals for $n$ = 1 and 4, excited by 375 nm light, at different ambient temperatures from -120 °C to 60 °C. By fitting the PL intensity dependence on temperature using $I(T) \propto 1/(1 + Ae^{-E_b/k_B T})$, we estimated exciton binding energy $E_b$. $E_b$ is the highest for 5IPA3-n1, reaching a value of 132 meV. $E_b$ decreases with increasing $n$, reaching a value of 94 meV for 5IPA3-n4 (Supplementary Fig. S3). By fitting the emission spectra near the band edge, we derived electronic temperature ($T_e$) and estimated Mott transition density for transparency. Considering the thickness of the organic spacer layer and an exciton Bohr radius of 0.6 nm adapted from $BA_2PbI_4$[41], the 3D Mott transition density for transparency exceeds $10^{19}$ $cm^{-3}$ (Supplementary Note 1 & Fig. S4).

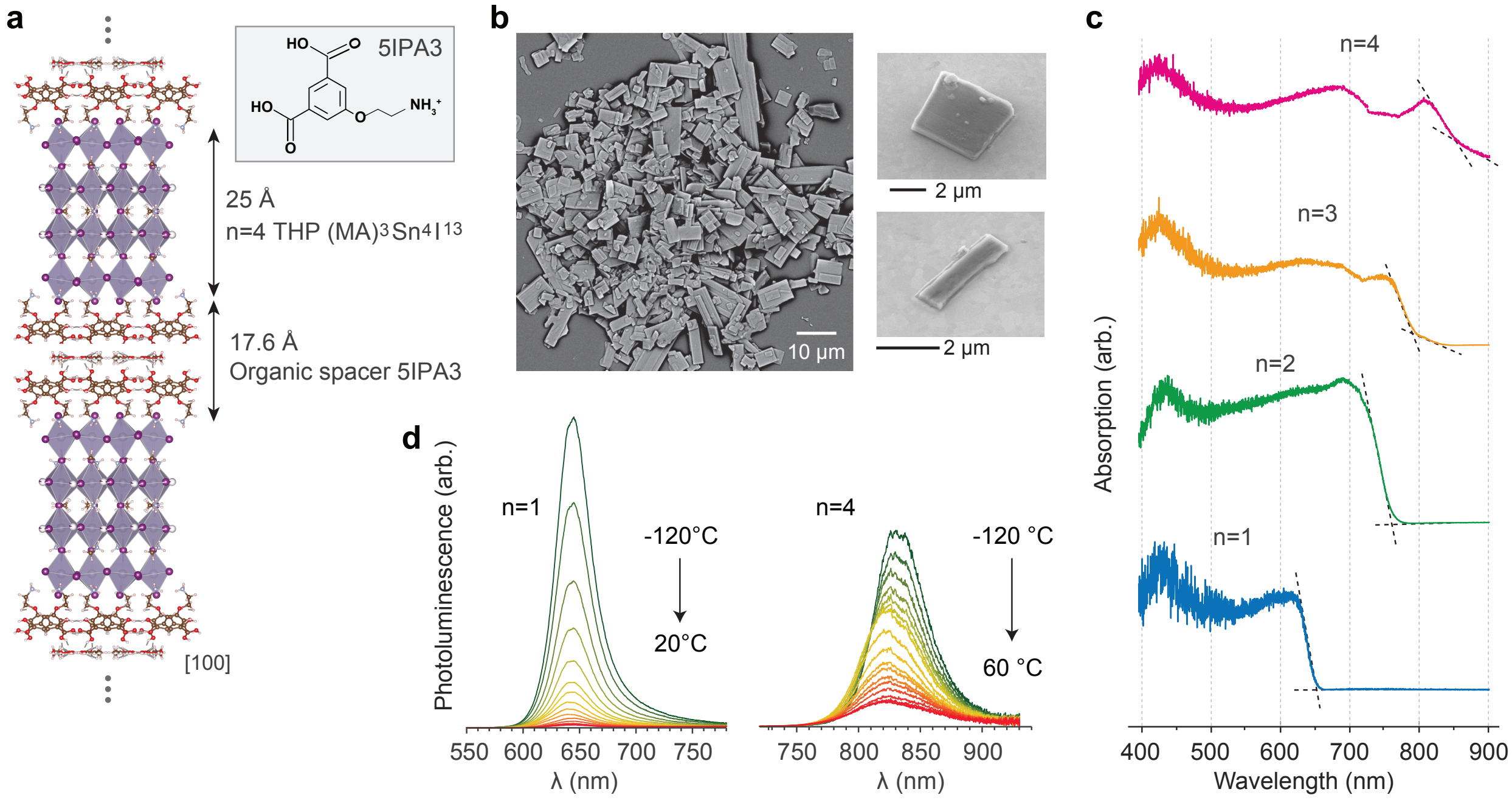

**Fig. 1. Material properties of layered tin iodide perovskite with 5IPA3 spacer. a**, Crystal structure of 5IPA3-n4 projected along [010]. The alternating organic and inorganic lattice have thicknesses of 17.6 and 24.0 Å, respectively. **b**, A representative scanning electron microscopy (SEM) image of abundant microcrystals produced by sonication. **c**, Diffusion reflectance spectra of $(5IPA3)_2(MA)_{n-1}Sn_nI_{3n+1}$ microcrystals with $n$ = 1 to 4. Dashed lines: linear fits to the absorption edge. **d**, Temperature-dependent photoluminescence spectroscopy of 5IPA3-n4 and 5IPA3-n1 microcrystals.

## Material stability at room temperature in air

In bulk $ASnI_3$ perovskites, where 'A' represents small organic cations such as methylammonium (MA) or formamidinium (FA), Sn(II) to Sn(IV) oxidation produces $SnO_2$, $SnI_4$, and AI.[42] $SnI_4$ and AI are then readily sublimed. Similar degradation route was verified in $n$=1 $PEA_2SnI_4$[43], as well as other quasi-2D systems. The growth and quality control of microcrystals have dramatical effects on their stability.[44] We measured a half-life of 200 minutes bulk $MASnI_3$ films in air using absorbance spectroscopy (Supplementary Fig. S5).

To assess the in-air stability of 5IPA3-n4 TIPs, we tracked the morphology and PL intensity evolution of both 5IPA3-n4 and $PEA_2SnI_4$ microcrystals (denoted as PEA-n1). The samples were stored in a dark condition between the brief optical measurements. While PEA-n1 microcrystals showed a rapid degradation, 5IPA3-n4 maintained stable PL for up to ~1000 minutes at room temperature in air (Fig. 2a). Some 5IPA3-n4 samples (#2 and #3 in Fig. 2a) exhibited a modest increase in PL intensity before it decays and is lost eventually. This is attributed to oxygen-induced defect healing, which is commonly observed in LHPs.[45]

Over time the original brown/yellow crystals gradually became transparent upon air exposure (Fig. 2b and Supplementary Fig. S6). This indicates complete Sn(II) oxidation, as the residual MAI and/or $SnO_2$ have minimal absorption of visible light. Oxidation proceeded from the edges and surfaces toward the center. A dark purple intermediate phase appeared during morphological evolution and disappeared upon full oxidation, often accompanied by a sudden PL drop. This dark phase might correspond to $SnI_4$ formation within the crystal interior before sublimation, potentially alongside $I^-$ oxidation to $I_2$, as suggested previously.[46] The dark phase emerged in just 20 minutes for PEA-n1 (Fig. 2b) but was significantly delayed for 5IPA3-n4 to 1-2 days (Fig. 2c).

Photostability was evaluated using continuous wave (CW) illumination at 375 nm with 1.3 W/cm$^2$. Figure 2d presented the PL time series of five individual microcrystals of 5IPA3-n4 and three individual microcrystals of PEA-n1. The mean lifetimes of the exponential PL decay were measured to be approximately 140 minutes for 5IPA3-n4 and 16 minutes for PEA-n1 (Supplementary Table S2). One 5IPA3-n4 sample showed a T90 (90% of initial intensity) of 18.3 minutes and a half-time (T50) of 3.5 hours.

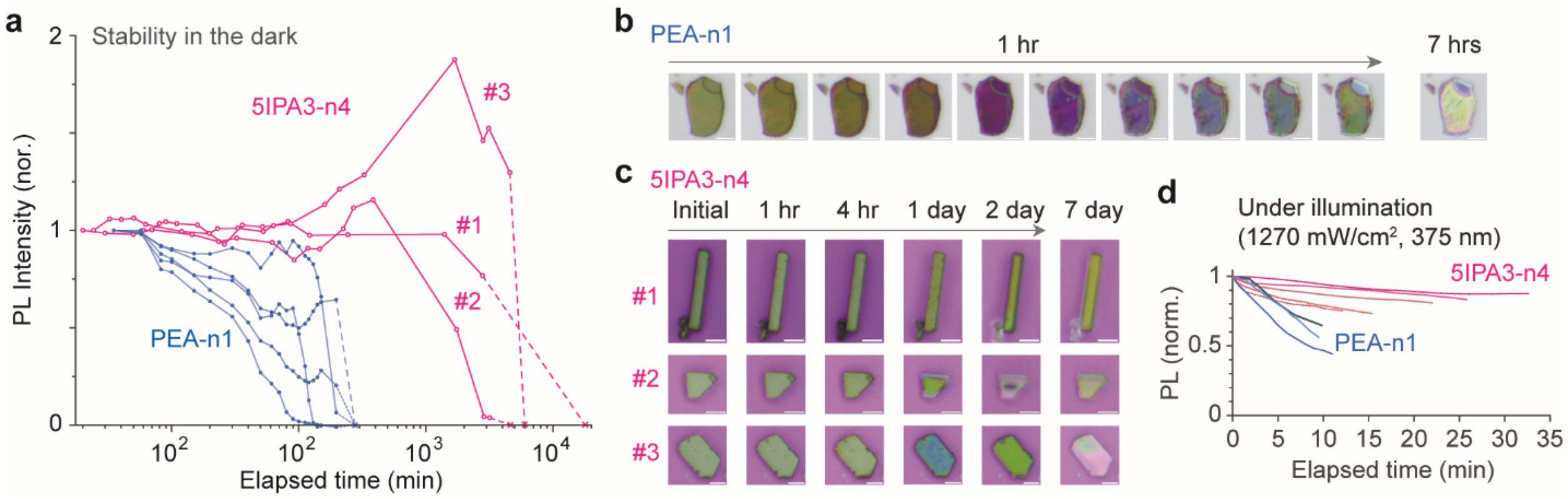

**Fig. 2. Stability of 5IPA3 TIPs in ambient conditions. a**, PL intensity variation of 5IPA3-n4 microcrystals (pink) and PEA-n1 TIPs (blue). Optical measurements were conducted using a CW laser at 385 nm with an intensity of 2.77 W/cm$^2$. **b**, Photos of a representative PEA-n1 microcrystal over time stored in the dark condition. **c**, Photos of the three 5IPA3-n4 samples in (a) over time. Scale bars: 5 µm. **d**, Photostability of 5IPA3-n4 (pink) and PEA-n1 (blue) microcrystals under continuous illumination of 375 nm light at an intensity of 1270 mW/cm$^2$ in ambient conditions.

**Lasing in air at room temperature**

We used a custom-built microscope to characterize lasing properties of microcrystals on a silicon substrate with a 200 nm thick oxide layer. The pump source was a frequency-doubled Q-switched Nd:YAG laser with a pulse width of 5 ns and an emission wavelength of 532 nm. The pump beam size was 30 μm to cover the entire area of microcrystals. Emission was collected using an objective lens with numerical aperture of 0.6, and the collected emission spectra was analyzed with a line-confocal spectrometer (Supplementary Fig. S7). A grating resolution of 0.1 nm was used for high resolution measurement, and resolution of 0.8 nm was for wideband measurement. All measurements were conducted in ambient air at room temperature.

We observed lasing action from 5IPA3 perovskites in the n=4 but not n < 4 phases (Fig. 3a). As the pump energy increased, narrowband, single-mode stimulated emission at 852 nm emerged on top of the broad spontaneous emission centered at 838 nm, with clamping occurring above the threshold pump energy of 1 mJ/cm$^2$ (Fig. 3b). The laser input-output curve (Fig. 3c) exhibited typical kink behavior, and fitting with a rate equation model yielded a spontaneous emission factor of $10^{-3}$. The linewidth decreased from 57 nm in spontaneous emission as lasing occurred (Fig. 3d). The narrowest linewidth, measured with high-resolution grating, was approximately 0.3 nm, corresponding to a lasing Q factor of 2850. Rapid material degradation occurred at a pump fluence of approximately 6 mJ/cm$^2$ (Supplementary Fig. S8).

The smallest lasing microcrystal was approximately 4 μm in its longest dimension. Finite-difference time-domain (FDTD) simulations suggest the presence of a high-Q dielectric resonance mode in the 4-μm-sized particle, with near-field profiles resembling whispering gallery modes in the square cavity, as shown in Fig. 2e. The refractive index of 5IPA3 (n=4) was estimated to be 2.3, based on the weighted average of the refractive indices of (MA)$SnI_3$ layers (index = 2.86)[47] and typical organic spacer layers (index = 1.5)[41]. The resonant mode had a cold cavity Q factor of 200, although experimental Q factors may be lower due to surface roughness, shape imperfections, and other factors. The calculated mode volume was 2 μm$^3$, which could be slightly overestimated due to the inherent field divergence of leaky cavity eigenmodes. The required gain coefficient for room-temperature lasing is approximately 800 cm$^{-1}$. Wide-field imaging of particles above the lasing threshold revealed interference patterns with bright four-side edges (Fig. 3f), consistent with the expected whispering gallery (WG) mode in a square cavity (Fig. 3g).

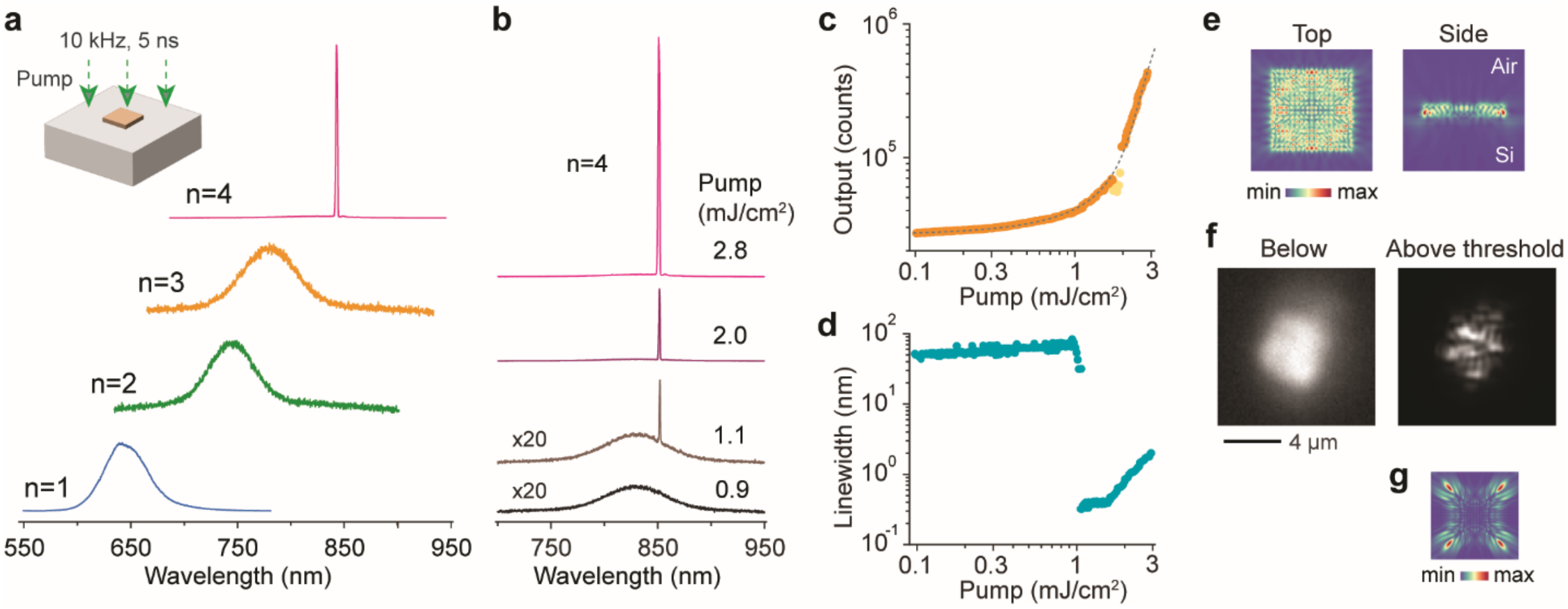

**Fig. 3. Lasing of 5IPA3-TIP microcrystals in ambient conditions. a**, (left top) Schematic of optically pumped 2D perovskite dielectric laser devices. (bottom) Emission spectra of microcrystals from n=1 to n=4 was recorded up to 3 mJ/cm². Only 5IPA3-n4 demonstrated lasing. **b**, Output spectra at pump fluences of 0.90, 1.11 2.04, and 2.84 mJ/cm². **c**, Measured light-in-light-out curve. Dashed line: theoretical fit. **d**, Measured linewidth at different pump fluences. **e**, FDTD-simulated electric field amplitude, |E|, of the lasing. **f**, Measured far-field emission profiles below and above lasing threshold. **g**, FDTD simulation for the far-field profile of the lasing mode.

## Plasmonic lasing in air at room temperature

To investigate potential plasmonic Purcell enhancement[48], we evaluated microcrystals placed on an ultrasmooth gold substrate (Supplementary Fig. S9). Figure 4a shows that 5IPA3 microcrystals on gold substrates exhibited lasing action from both n=3 and n=4 microcrystals, with a lasing threshold ranging from 0.75 to 1.2 mJ/cm² and a spontaneous emission factor of $10^{-2}$. The narrowest observed linewidth was ~0.35 nm, corresponding to a lasing Q-factor of 2440. The laser spectra, input-output intensities, linewidth narrowing, and spatial emission patterns for representative devices are presented in Fig. 4 and Supplementary Figs. S10 and S11. FDTD simulation showed a hybrid plasmonic-photonic mode with a cold cavity Q-factor of 50 and a mode volume of $3\times10^{-19}$m³, yielding a Purcell factor of 2 (Supplementary Fig. S10). The intrinsic gain required for room-temperature lasing, considering the Purcell enhancement, is ~1690 $cm^{-1}$. Plasmon-exciton coupling was evidenced by the appearance of wiggling patterns and spatial inhomogeneity of Purcell enhancement in the confocal fluorescence detection of n=1 microcrystals near the plasmonic dispersion asymptote (Fig. 4a and Supplementary Fig. S12).[48]

Transient lifetime spectroscopy (120 ps resolution) revealed an accelerated lifetime in a double-exponential fit: 90.4% of 240 ps and 9.6% of 610 ps for microparticles on gold substrates, compared to 86.5% of 410 ps and 13.5% of 1.4 ns for microparticles on thermal-oxide silicon substrates (Fig. 4f). Assuming that the fastest channel corresponds to nonradiative decay and the slower one to radiative recombination, recombination on gold occurred approximately twice as fast, consistent with FDTD calculations.

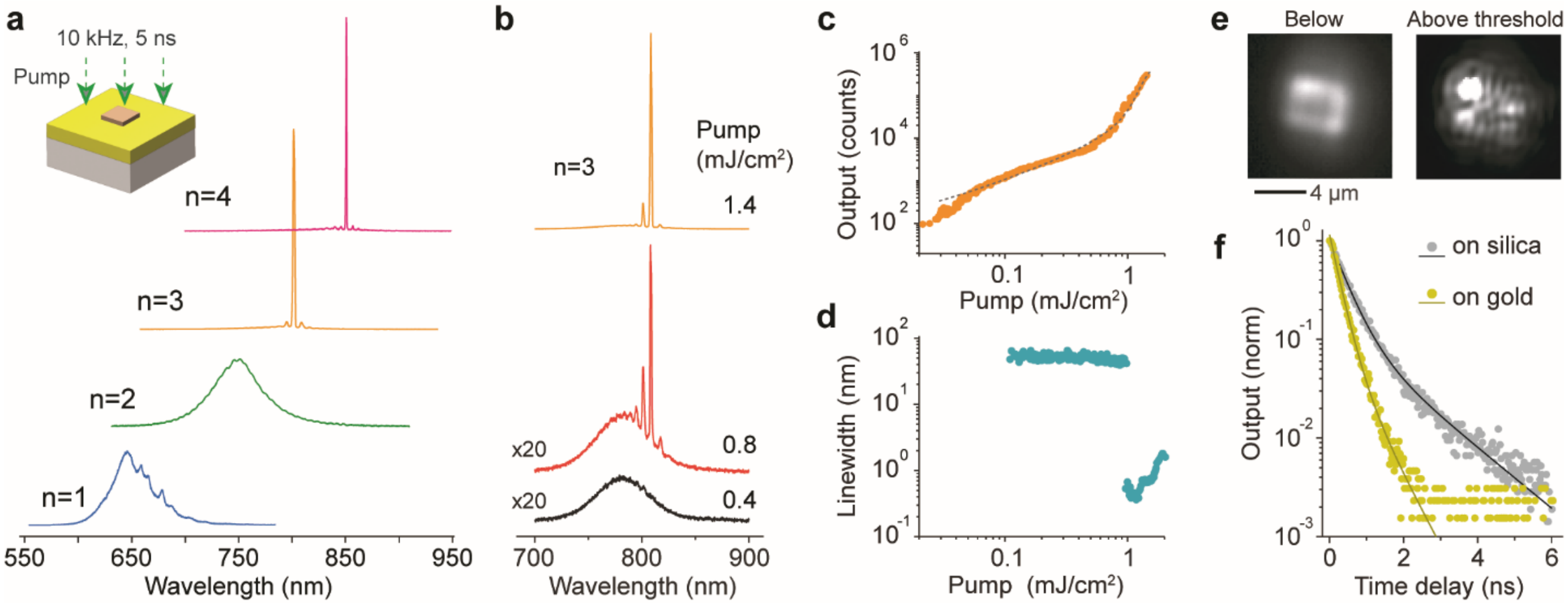

**Fig. 4. Plasmonic lasing in 5IPA3-TIPs in ambient conditions. a**, (left top) Schematic of an optically pumped 2D perovskite sample on a gold substrate. (bottom) Typical emission spectra from microcrystals with n=1 to n=4 phases, at pump fluences of ~ 3 mJ/cm². 5IPA3-n3 and -n4 samples demonstrated lasing. **b**, Output spectra from an 5IPA3-n3 microcrystal at pump fluences of 0.42, 0.81, 1.1, and 1.4 mJ/cm². **c**, Measured light-in-light-out curve. Dashed line: theoretical fit. **d**, Measured linewidth at different pump fluences. **e**, Measured far-field emission profiles below and above lasing threshold. **f**, Transient lifetime decay curves below threshold from 5IPA3-n4 TIP microcrystals on gold (yellow) and thermal oxide silicon (gray) substrates.

### Pump effect on laser stability

We investigated picosecond pumping at 765 nm, near the band edge of n=4 microcrystals on gold substrates, with a 2.5 MHz repetition rate (Fig. 5a). The typical lasing threshold was 0.22 mJ/cm² per pulse, about four times lower than that for nanosecond pumping (Fig. 5b). The spontaneous emission factor was approximately 0.001. The linewidth at 1.5 times the threshold was 0.59 nm (Fig. 5c), and interference patterns were observed above the threshold (Fig. 5d)

Next, we compared the laser emission stability of 5IPA3-n4 perovskites on a gold substrate under nanosecond and picosecond pumping. In a sample under nanosecond pumping, mode hopping occurred after ~$10^4$ shots, followed by a blue shift of 3 nm and a 90% intensity drop after ~$3\times10^4$ shots (Fig. 5e).

Other samples exhibited a 1.83 nm blue shift with a 95% intensity drop (Supplementary Fig. S13). In contrast, under picosecond pumping, laser emission persisted for a half billion pump pulses (Fig. 5f and Supplementary Fig. S13). Figure 5g shows decay curves on a log scale, showing rapid degradation after $10^8$ pulses.

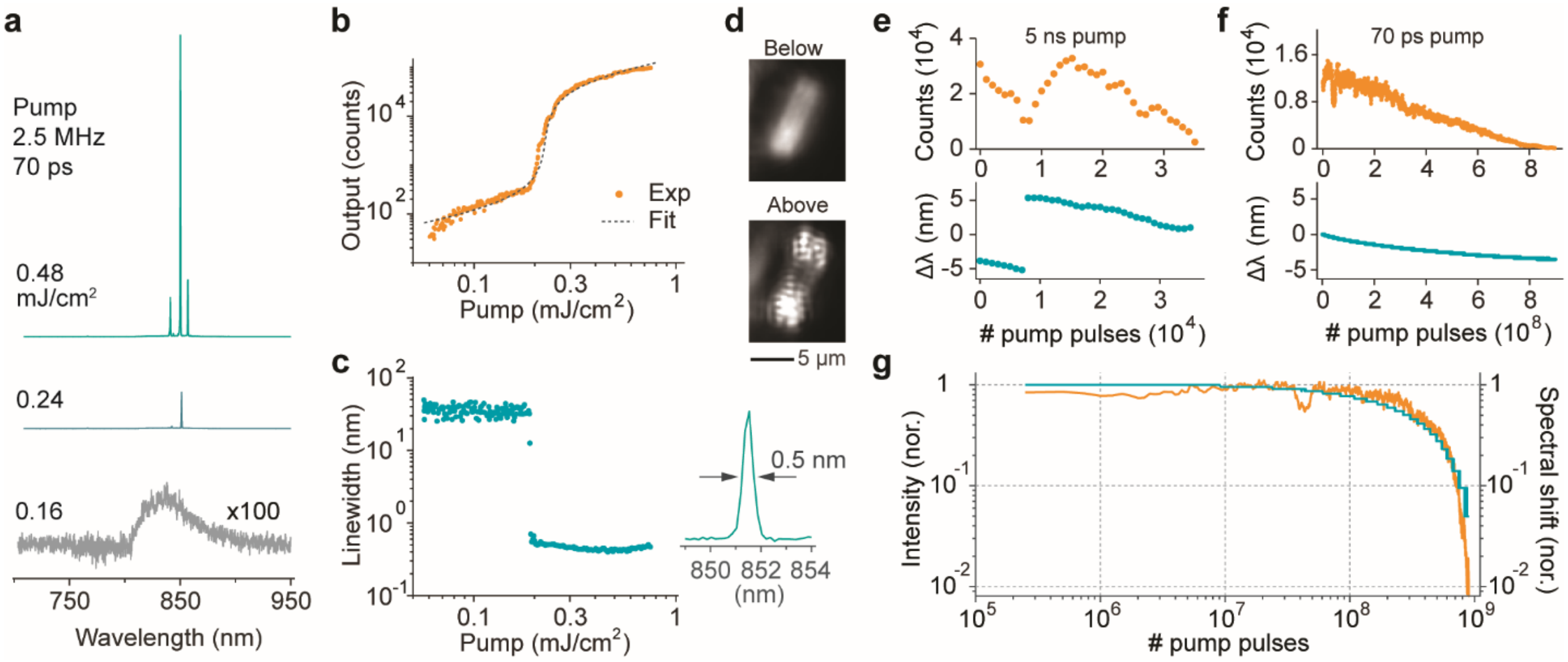

**Fig. 5. Laser stability in ambient conditions. a**, Output spectra of 5IPA3-n4 microcrystals on a gold substrate under picosecond pumping at various pump fluences: 0.16, 0.24, 0.36, and 0.48 mJ/cm². **b**, Measured light-in-light-out curve. Dashed line: theoretical fit. **c,** Measured linewidth at different pump fluences. (right) Emission spectra at 1.5 times the lasing threshold are shown. **d**, Measured far-field emission profiles below and above lasing threshold. **e**, The intensity (top) and wavelength (bottom) variations of an n=4 plasmonic device under nanosecond pumping. **f**, Decay curves of another device under picosecond pumping at 1.5 times the lasing threshold. **g**, Aging curves on a log scale for intensity (orange) and wavelength (green).

## Discussion

The incorporation of the 5IPA3-based hydrogen-bonding network significantly enhanced the photostability of 2D and quasi-2D TIPs. This improved stability enabled the first demonstration of room-temperature lasing in air from 5IPA-n3 and 5IPA-n4 microcrystals. Previously, lasing from n=1 tin-iodide 2D perovskite nanowires with similar hydrogen-bonding organic sublattices[28] was achieved only at 88 K under inert atmosphere. Room-temperature lasing was observed from n=2 quasi-2D TIP exfoliated flakes in nitrogen, with a linewidth of ~2 nm and a threshold of around 0.4 mJ/cm².[27] All of these prior demonstrations relied on inert gas protection. Furthermore, these earlier systems required threshold peak power on the order of GW/cm², much higher than compared the kW to MW/cm² thresholds typical for 3D $CsPbBr_3$ and $MAPbI_3$ microlasers. In contrast, our 5IPA-based quasi-2D TIPs achieved lasing in air with a lower threshold of ~200 kW/cm$^2$ (Supplementary Table S3 and Fig. S14). This result highlights a major step toward practical device operation. In addition, we reported the first plasmonic lasing in TIPs. The optical gain coefficients of 5IPA3-n3 and 5IPA3-n4 TIPs are estimated to exceed 1000 cm$^{-1}$. We hypothesize that longer exciton diffusion lengths and enhanced intralayer energy transport in high-n TIPs contribute to a stronger Purcell effect near gold substrates for hybrid plasmonic-photonic modes. This work provides a foundation for future efforts in spacer engineering and material optimization toward practical photonic and optoelectronic applications.

## Author Contributions

S.C., W.S. L.D., and S.-H.Y. designed the study. W.S. and J.K. synthesized the materials. W.S. characterized material properties. S.C. conducted the lasing experiments and modeling. S.C., W.S., L.D., and S. H.Y. wrote the manuscript.

## Acknowledgements

This research was supported in part by funding from the National Institutes of Health (grants R01EB033155 and R01EB034687) and from US Department of Energy, Office of Basic Energy Sciences, under award no. DE-SC0022082.

## Conflict of Interest

None.

Supplementary Information for

# Air-Stable Room-Temperature Quasi-2D Tin Iodide Perovskite Microlasers

Sangyeon Cho[1,2†], Wenhao Shao[3,4†], Jeong Hui Kim[3], Letian Dou[3,5,6*], Seok-Hyun Yun[1,2,7*]

[1]Wellman Center for Photomedicine, Massachusetts General Hospital, 65 Landsdowne St., Cambridge, MA 02139, USA

[2]Harvard Medical School, Boston, MA 02139, USA

[3]Davidson School of Chemical Engineering, Purdue University, West Lafayette, IN, 47907 USA

[4]Department of Chemistry, University of Georgia, Athens, GA 30602 USA

[5]Department of Chemistry, Purdue University, West Lafayette, IN, 47907 USA

[6]Department of Chemistry, Emory University; Atlanta, GA 30322, USA.

[7]Harvard-MIT Health Sciences and Technology, Cambridge, MA 02139, USA

[†]Equal Contribution

*Correspondence to: dou10@purdue.edu or syun@hms.harvard.edu

## MATERIALS AND METHODS

### Materials

Tin iodide ($SnI_2$, AnhydroBeads™, −10 mesh, 99.99% trace metals basis, #409308), hydroiodic acid (HI, 57 wt.% in $H_2O$), hypophosphorous acid ($H_3PO_2$, 50 wt.% in $H_2O$), and acetic acid (AcOH) were purchased from Millipore Sigma. Methylammonium iodide (MAI, >99.99%, #SKU MS101000) were purchased from Greatcell Solar Materials. 2-(3,5-dicarboxyphenoxy)ethan-1-aminium iodide (5IPA3I) was synthesized.

### Layered perovskite microcrystal growth

Crystals were grown with a standard fast recrystallization method from aqueous solvent. In detail, perovskite solid precursors (5IPA3I, $SnI_2$, and MAI) were weighed in a $N_2$-filled dry glovebox into a 2 mL scintillation vial equipped with a PTFE membrane cap. The vial was then capped transferred into a $N_2$-filled wet glovebox where aqueous solvents (HI, $H_3PO_2$, and acetic acid) were added. The detailed precursor composition was summarized in Supplementary table S1. The capped vial was then heated on a hotplate at 140-160 °C (surface temperature) until a clear and refluxing solution was obtained. Then, the hotplate was turned off, and the vial was left cooled on the hotplate naturally and undisturbed until the temperature had stabilized at room temperature. Dark-red (n=1) or black (n>1) crystals precipitated during this process.

The bulk crystals were sonicated in the mother solution to generate abundant microcrystals. The fully stabilized vial was transferred out from the glovebox and sonicated in a bath sonicator in air for 3 minutes. The vial was well-capped during this period, yet elongated sonication would inevitably accelerate the sonication of the solution, which should be avoided. The vial was transferred back to the wet glovebox.

Microcrystals grown with this method were then transferred with the solution to a glass substrate. The crystals were picked up by gently touching the droplet surface with polydimethylsiloxane (PDMS) stamps (from GelPak). Leftover solvents were gently absorbed with a piece of filter paper.

**Supplementary Table S1. Precursor and solvent composition for layered perovskite microcrystal growth.**

| *n* | $SnI_2$ (mg) | MAI (mg) | 5IPA3I (mg) | HI (μL) | $H_3PO_2$ (μL) | AcOH (μL) |
|---|---|---|---|---|---|---|
| 1 | 20 | / | 2 | 50 | 25 | 37.5 |
| 2 | 5 | 1.5 | 2 | 50 | 25 | 37.5 |
| 3 | 10 | 1.5 | 2 | 50 | 25 | 37.5 |
| 4 | 10 | 7.5 | 1 | 66.6 | 33.4 | 50 |

### $MASnI_3$ film fabrication and stability in air

20 $\mu$L $MASnI_3$ solution (0.1 M, weighing and preparation in a $N_2$-filled glovebox) was dropped on a UV-ozone cleaned glass slide (1.5*1.5 cm) and was spin coated at 4000 rpm for 2 minutes, followed by annealing at 100 °C for 10 minutes. UV-vis absorption of the film was measured continuously at 20 °C in air and in dark conditions. The absolute absorbance was corrected with an empty glass slide as the background.

### Steady state photoluminescence (PL)

Steady state PL spectra were measured with an Olympus BX53 microscope under epi-detection reflection mode either with an integrated X-CITE 120Q Mercury lamp or a 375 nm continuous wave laser. The excitation light and power density used were indicated in each related figure inset. The filter set contains a 385 nm bandpass filter and a 420 nm long-pass filter. The collected PL signals were analyzed by a spectrometer (SpectraPro HRS-300). Spectra were intensity-calibrated using a NIST traceble StellarNet SL1-CAL VIS-NIR Tungsten Halogen lamp to generate the standard curve. PL time series (Figure 2a and 2c) were assessed in air while the temperature-dependent PL (Fig. 1d and Supplementary Fig. S3) was assessed in a $N_2$-filled heating stage (Instec), both with the microcrystals deposited on $Si/SiO_2$ substrates.

**Scanning electron microscope (SEM) imaging.**

Figure 1b was collected with a Teneo Volumescope FEG SEM with a T1 detector at 5.00 kV and 10.0 spot size. The individual crystals in Figure 1b and S9 were collected by using Zeiss Gemini 360 FESEM with a SE detector at 5.00 kV. Figure S1 was collected with a FEI Nova NanoSEM Field Emission SEM with a Everhart–Thornley detector (ETD) at 5.00 kV and 4.5 spot size.

**Laser experiments.** For the laser experiment, the specimen is placed in a home-built epi-fluorescence microscopy setup. The pump source is a frequency-doubled NDYAG laser (532 nm), with a repetition rate of 1 kHz and a pulse duration of 5 ns or an amplified frequency doubled fiber laser emitting at 765 nm with a repetition rate of 2.5 MHz and a pulse duration of 70 ps. Using a 0.6 NA, 50× air objective lens, the full-width-at-half-maxima (fwhm) size of the pump beam on the sample is about 20 μm. The system utilized either a 0.6 NA, 50x air objective lens or a 0.4 NA, 20x air objective lens. Emission from the sample, collected by the objective lens, passed through a dichroic mirror and a dichroic f ilter before being split into two paths. One path was directed to a silicon-based EMCCD camera (Luca, Andor) for wide-field imaging. The other path was directed to an EMCCD (Shamrock, Andor) in a spectrometer equipped with two gratings: Grating 1 (300 lines/mm, 500 nm blaze) with a 100 μm slit (resolution: 0.7-0.9 nm) and Grating 2 (1200 lines/mm, 500 nm blaze) with a 100 μm slit (resolution: 0.13 nm). The second output port of the spectrometer was coupled with an avalanche photodiode (APD) and a time-correlated single photon counting (TCSPC) system (Timeharp 260, PicoQuant) to perform transient lifetime spectroscopy. The instrument response function, characterized by using a picosecond pump laser, exhibited a time resolution of 120 ps.

**Numerical simulation.** Finite difference time domain (FDTD) simulations were performed using commercial software (Lumerical). Mie scattering simulation employed a total-field scattered-field plane-wave source, while dipole simulations utilized both electric and magnetic dipoles placed inside semiconductor particles. Time-dependent electric and magnetic fields were recorded using densely positioned point-like time monitors within the semiconductor particles. Resonance frequencies ($\varpi$) and the full-width-half-maximum spectral widths ($\Delta\varpi$) were used to calculate the quality factors (Q) of low-Q modes: $Q = \varpi / \Delta\varpi$ . For high Q modes that did not decay completely within the simulation timeframe, Q values were determined from the slope of the electric field decay profiles. Three-dimensional near-field field patterns were obtained using an array of two-dimensional field monitors, while far-field emission patterns were calculated using a box monitor that encompassed the particle.

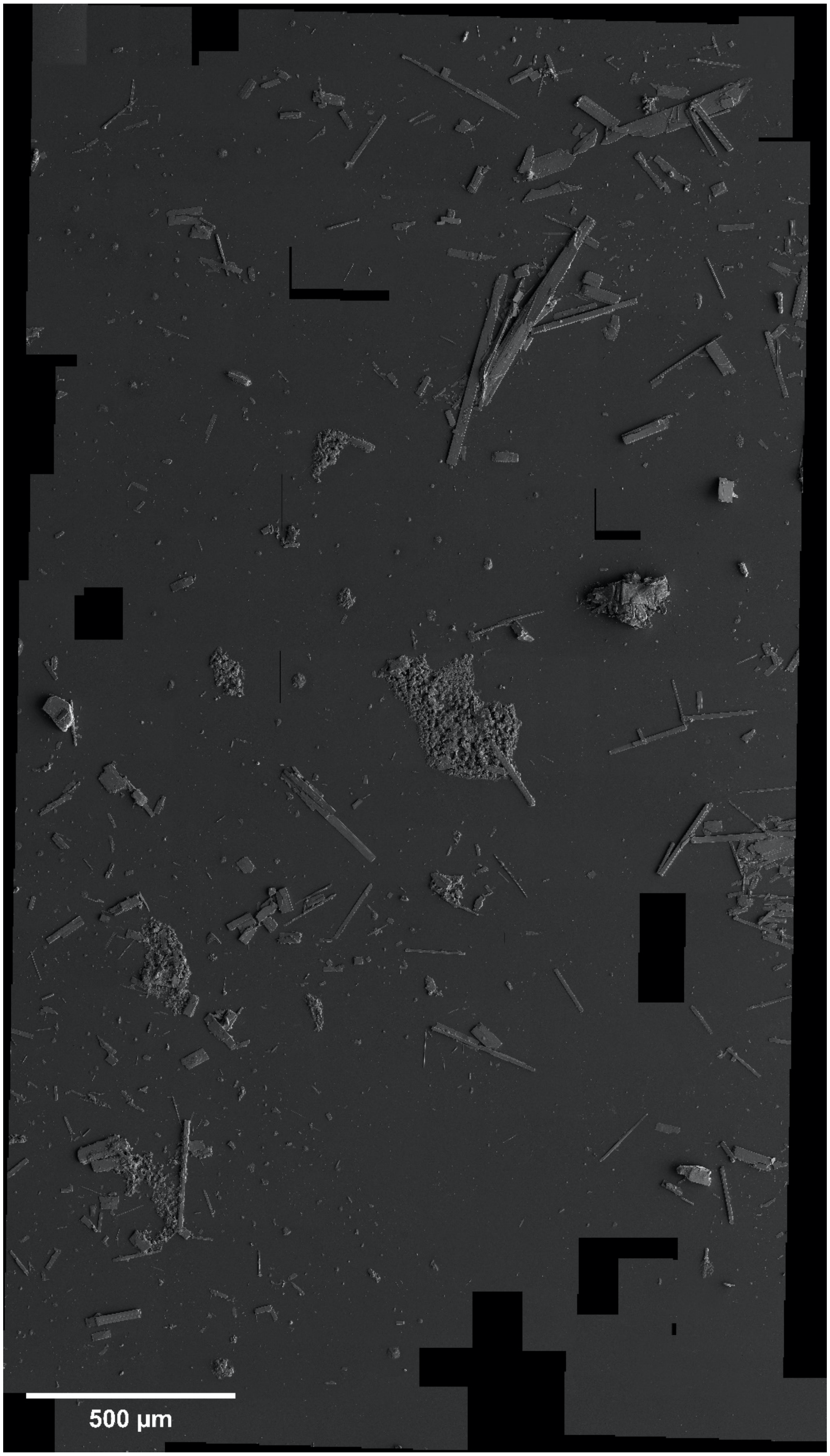

**Supplementary Fig. S1. Overall abundance of 5IPA3-n4 microcrystals under SEM.** Microcrystals were deposited on a PDMS stamp for SEM. This is the actual image used for demographic analysis in Fig. S2. SEM specifics: FEI Nova; Everhart–Thornley detector (ETD); 5.0 kV; spot size 4.0.

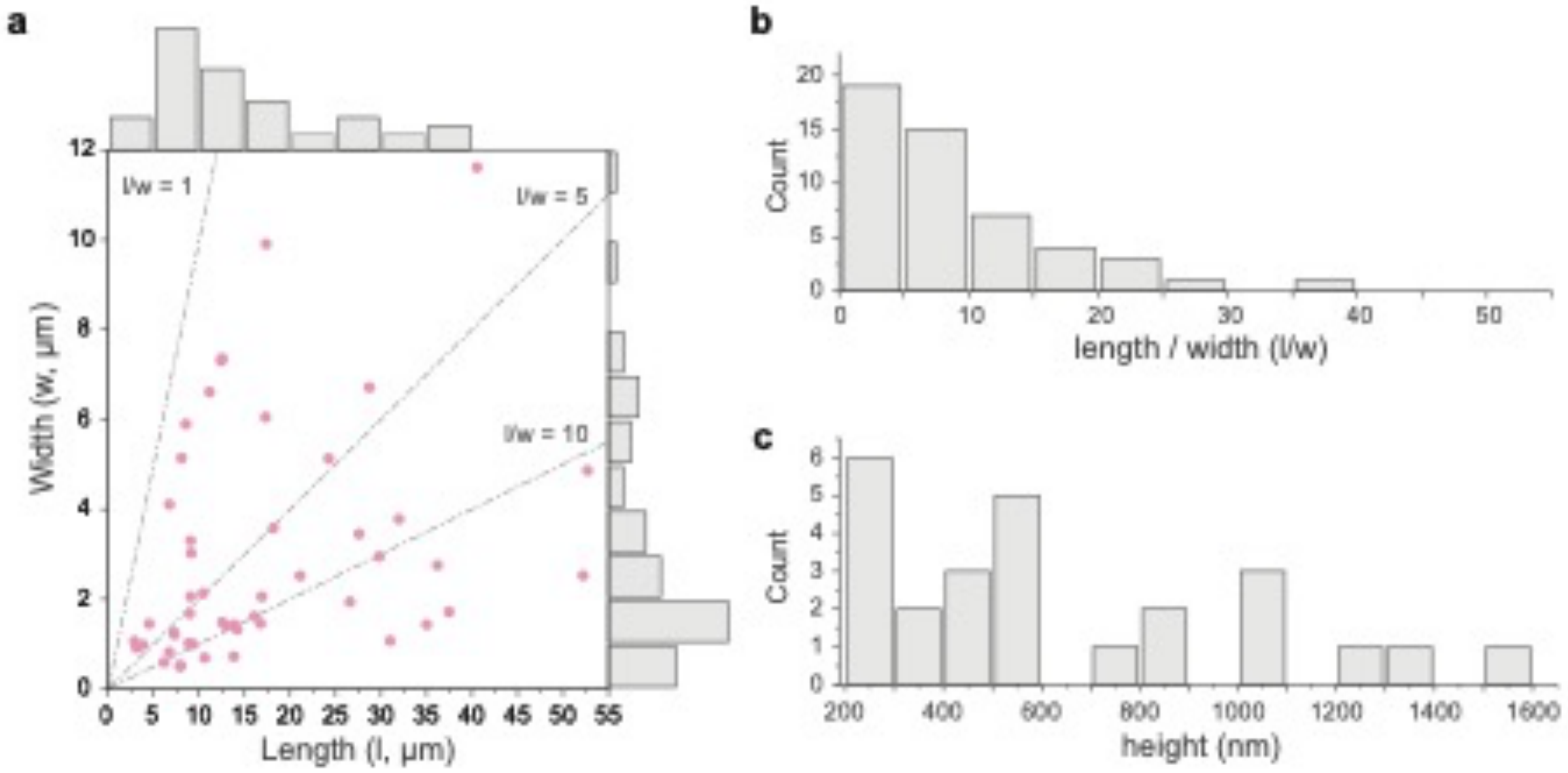

**Supplementary Fig. S2. Quantified dimensions of 5IPA3-n4 microcrystals. a,** Length (*l*) and width (*w*) distribution over a population of 50 crystals from the large-area SEM image shown in Fig. S3. Aspect ratio (*l/w*) guidelines are included. **b,** *l/w* aspect ratio distribution. **c,** height (*h*) distribution over a population of 25 crystals investigated by tapping-mode atomic force microscope (AFM).

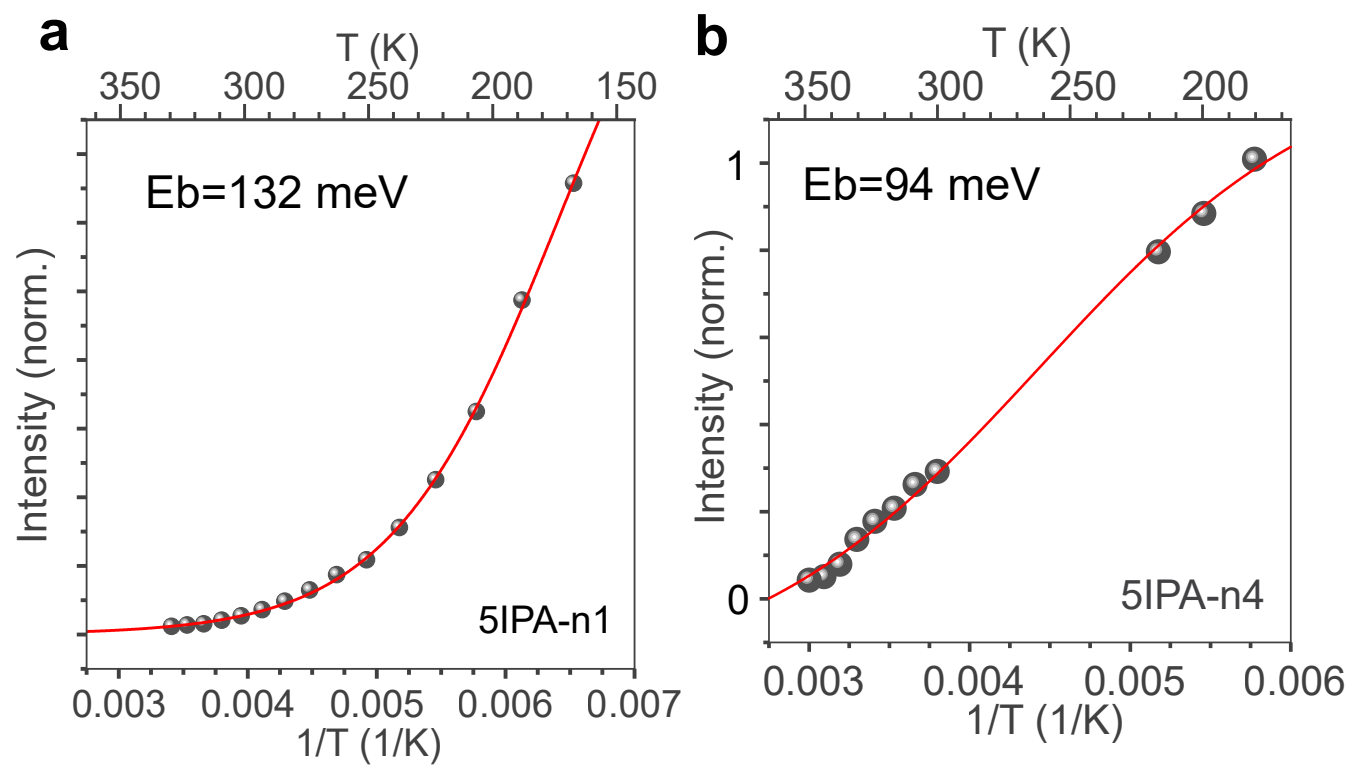

**Supplementary Fig. S3. Temperature-dependent photoluminescence spectroscopy of n=1 and n=4 $5IPA_3$ microcrystals. (a-b)** The integrated PL intensity at different temperatures for **a,** 5IPA-n1 **b,** 5IPA-n4 microcrystals. The effective exciton binding energy of the perovskite is extracted using the Arrhenius formula fitting $I(T) = I_0/(1 + Ae^{-\frac{E_b}{k_b T}})$ (red line). The corresponding PL emission spectra for 5IPA-n1 and 5IPA-n4 microcrystals are presented in Fig. 1d.

**Supplementary Note 1.** Layer number (n) dependence on Mott carrier density

2D tin halide perovskites with varying layer numbers (n) exhibit exciton binding energies that exceed the thermal energy at room temperature (~25 meV). At low pump fluence, photoexcited carriers primarily form bound excitons. As the pump fluence increases, these excitons begin to dissociate into free carriers. When the carrier density surpasses a critical threshold—known as the Mott transition, which occurs when the interparticle separation $r \propto n^{-1/3}$ approaches the Debye screening length —an electron-hole plasma (EHP) is formed, and optical gain begins to build up due to band gap renormalization.[1] Therefore, the Mott transition density can be interpreted as the transparency carrier density. Previously, the formation of electron-hole plasma (EHP) has been observed in various 3D perovskite systems and bulk-like nanocrystals which do not exhibit strong quantum confinement[2,3].

3D Mott transition density can be written as $\rho_{Mott} \approx 0.028(k_B T_e/(E_B a_B{}^3)$, where $k_B$ is Boltzmann constant, $T_e$ (= 709 K; Supplementary Fig. S4a) is electronic temperature, $E_B$ is exciton binding energy, and $a_B$ is Bohr radius.[4] Since the Bohr radius is inversely proportional to the square root of the exciton binding energy ($a_B \propto 1/\sqrt{E_B}$), the Mott carrier density scales as the square root of the binding energy ($\rho_{Mott} \propto \sqrt{E_B}$). Due to the thickness of the organic spacer layer, exciton transport exhibits quasi-3D behavior. As a result, at high carrier densities, the lasing behavior resembles that of bulk 3D systems. According to previous studies[5] on quasi-2D lead halide perovskites with butylammonium (BA) organic spacers, the Mott carrier density exhibits a clear dependence on the layer number n, as shown in Supplementary Fig.S4b. As n increases from 1 to 4, the exciton binding energy decreases from approximately 467 meV to 157 meV, while the Bohr radius increases from about 0.6 nm to 1 nm.[5] Correspondingly, the estimated Mott carrier density drops from ~2x$10^{19}$ $cm^{-3}$ to ~9x$10^{18}$ $cm^{-3}$. For 5IPA3-based 2D tin iodide perovskites, the exciton binding energies were estimated to be 132 meV and 94 meV Supplementary Fig.S3) and assuming similar Bohar radius, corresponding to Mott densities of approximately ~$10^{19}$ $cm^{-3}$ to ~7x$10^{18}$ $cm^{-3}$. These values are about an order of magnitude higher than that of 3D $CsPbBr_3$ perovskites[6], but they remain within the achievable range for lasing applications.

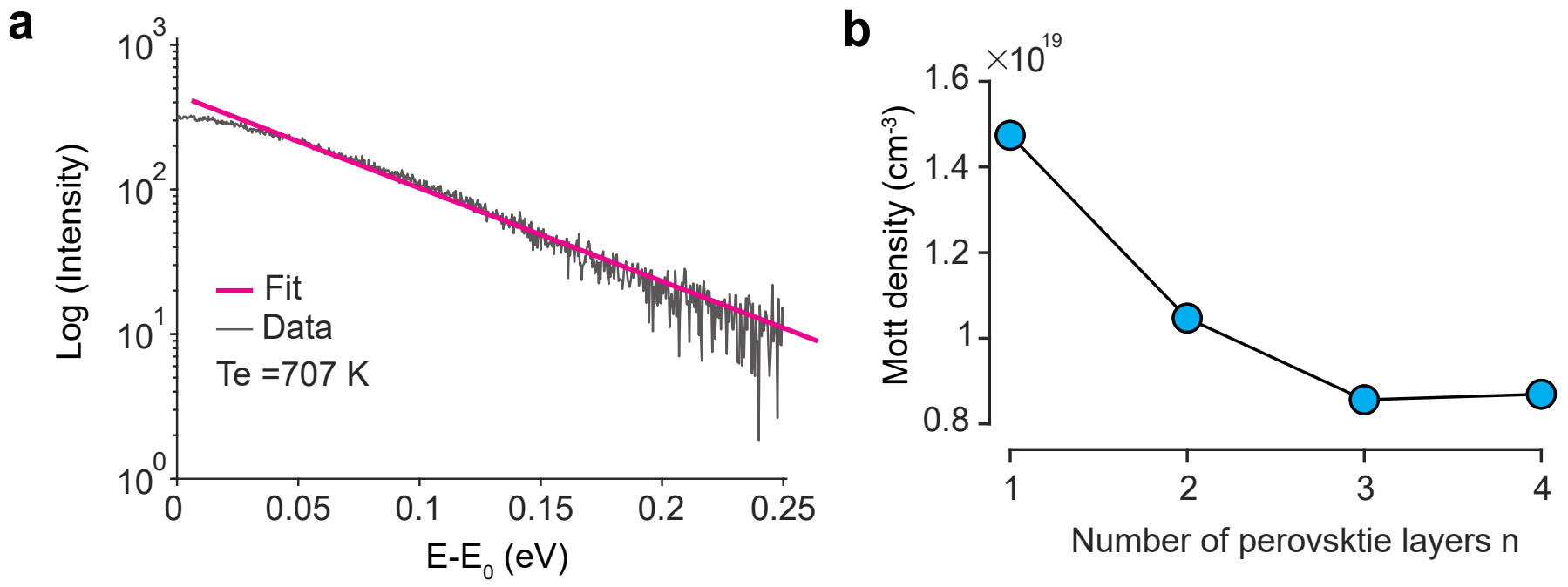

**Supplementary Fig. S4. a,** Extraction of $T_e$ for the estimating of $\rho_{Mott}$. The blue-side tail of the photoluminescence (PL) spectra at P=0.9Pth is shown. The tail was fitted using the classical Maxwell-Boltzmann distribution function, which characterizes the electron-hole plasma (EHP) regime: $N(E) \propto \exp\left(-(E - E_0)/k_B T_e\right)$, where $N(E)$(proportional to PL intensity) represents the carrier population at energy $E$, $E_0$ is the peak emission energy, and $T_e$ is the electronic temperature extracted from the fit. **b,** Estimated Mott carrier densities of quasi-2D BA-based lead halide perovskites across different layer numbers. Exciton binding energies and Bohr radii are adapted from previously published data by Blancon, J.-C. et al., Scaling law for excitons in 2D perovskite quantum wells, Nature Communications 9, 2254 (2018).

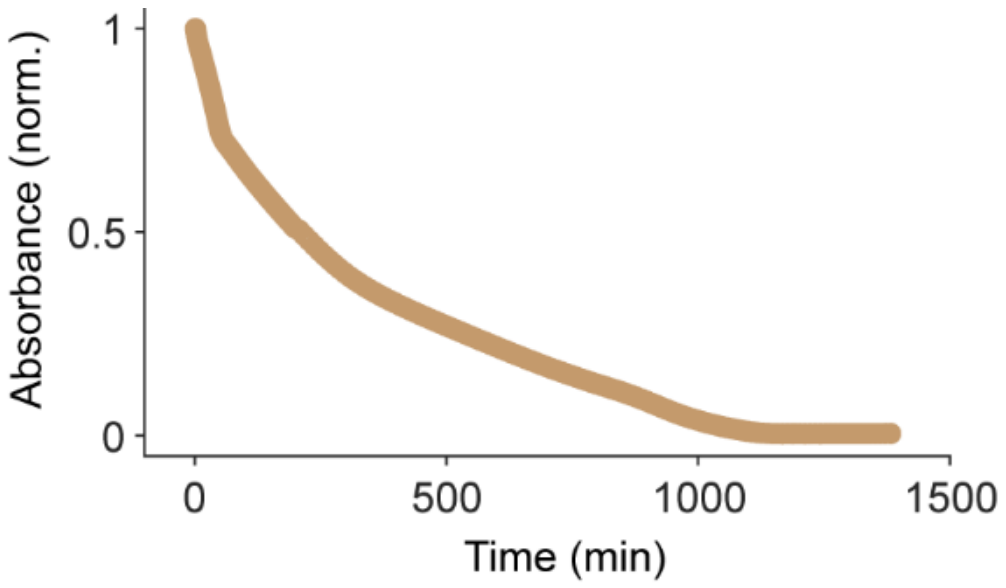

**Supplementary Fig. S5.** Air-stability of MA-based 3D TIP films evaluated with their integrated absorption from 400-1000 nm.

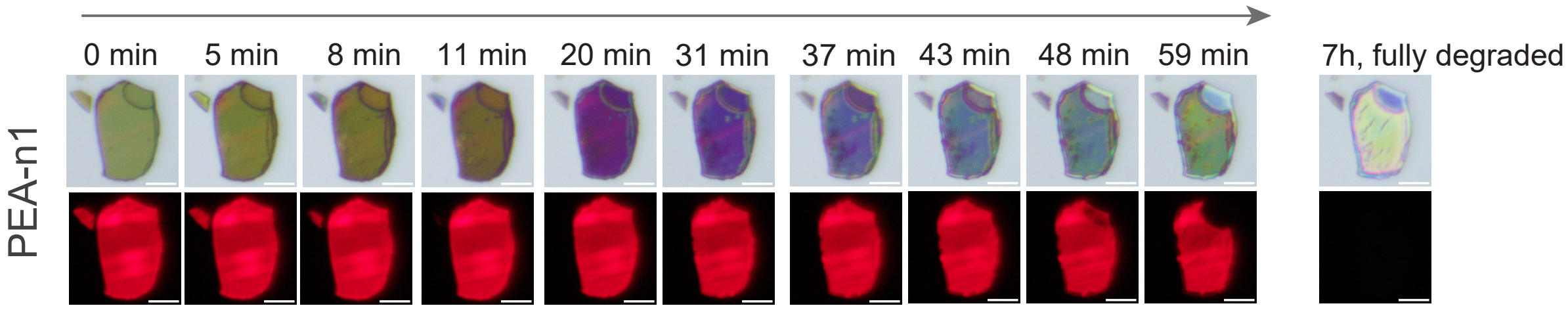

**Supplementary Fig. S6.** Morphology and fluorescence evolution of a representative PEA-n1 microcrystal from Fig. 2a at each time point during air exposure. Scale bars: 5 μm.

| | T90 (s) | $A_1$ | $\tau_1$ (s) | $A_2$ | $\tau_2$ (s) | $\tau_{avg}$ (s) |
|---|---|---|---|---|---|---|
| 5IPA3-n4<br>#1 | 1098 | 1.06E+00 | 1.24E+04 | | | 1.24E+04 |
| #2 | 157 | 8.58E-02 | 1.12E+02 | 9.05E-01 | 4.36E+03 | 3.99E+03 |
| #3 | 76 | 1.16E-01 | 6.01E+01 | 8.75E-01 | 4.40E+03 | 3.89E+03 |
| #4 | 356 | 7.94E-02 | 1.71E+02 | 9.20E-01 | 1.09E+04 | 1.00E+04 |
| #5 | 922 | 3.17E-02 | 2.58E+01 | 9.69E-01 | 1.14E+04 | 1.11E+04 |
| PEA-n1<br>#1 | 207 | 1.10E+00 | 1.09E+03 | | | 1.09E+03 |
| #2 | 81 | 9.97E-01 | 7.58E+02 | | | 7.58E+02 |
| #3 | 182 | 1.10E+00 | 9.81E+02 | | | 9.81E+02 |

**Supplementary Table S2. Photostability of 2D microcrystals.** The time-dependent photoluminescence intensity $I(t)$ obtained under continuous wave excitation at 375 nm and 1266 mW/cm² was analyzed using a double exponential fitting model assuming $\lim_{t\to\infty} I(t) = 0$.

$$I(t) = A_1 e^{-\frac{t}{\tau_1}} + A_2 e^{-\frac{t}{\tau_2}}$$

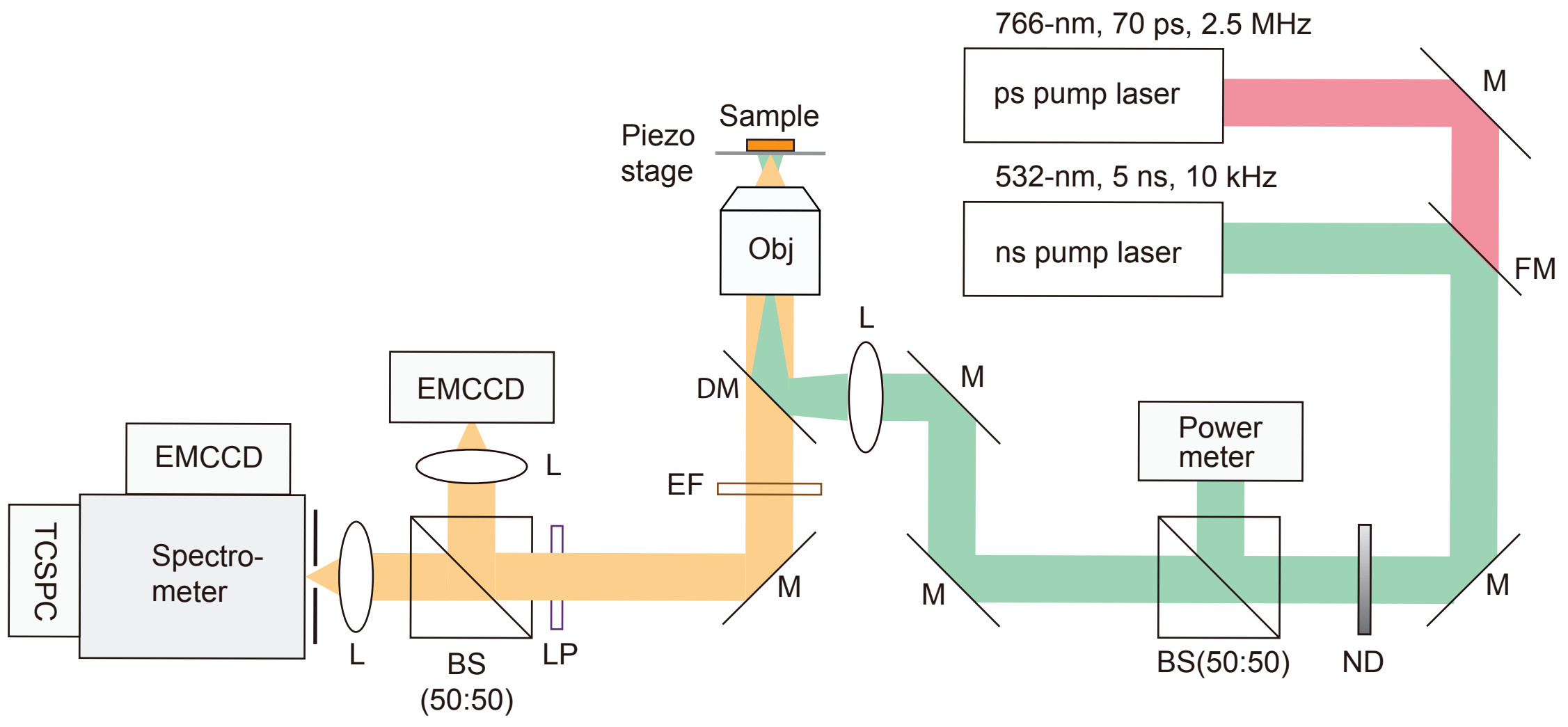

**Supplementary Fig. S7. Laser optical characterization setup.**

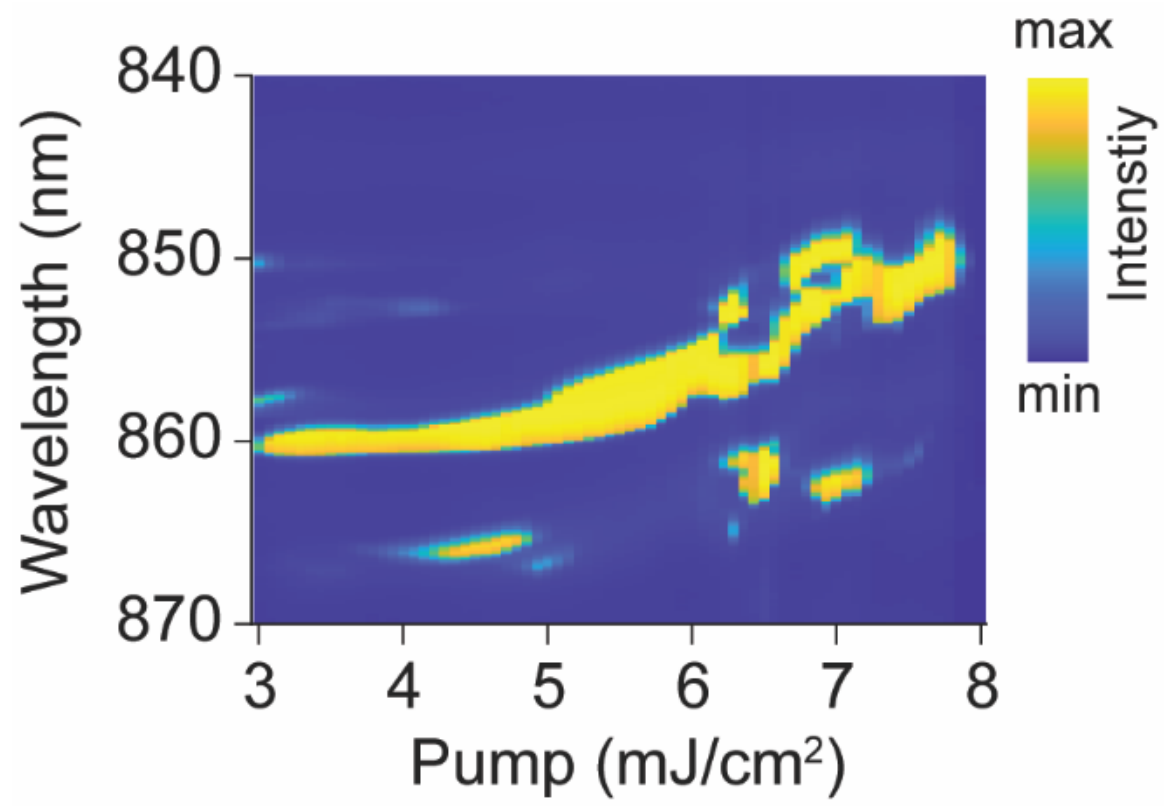

**Supplementary Fig. S8.** The damage threshold of 5IPA3 perovskite (n = 4) was characterized by varying nanosecond pump energies. Unstable lasing behavior begins to appear at fluences above approximately 6 mJ/cm$^2$.

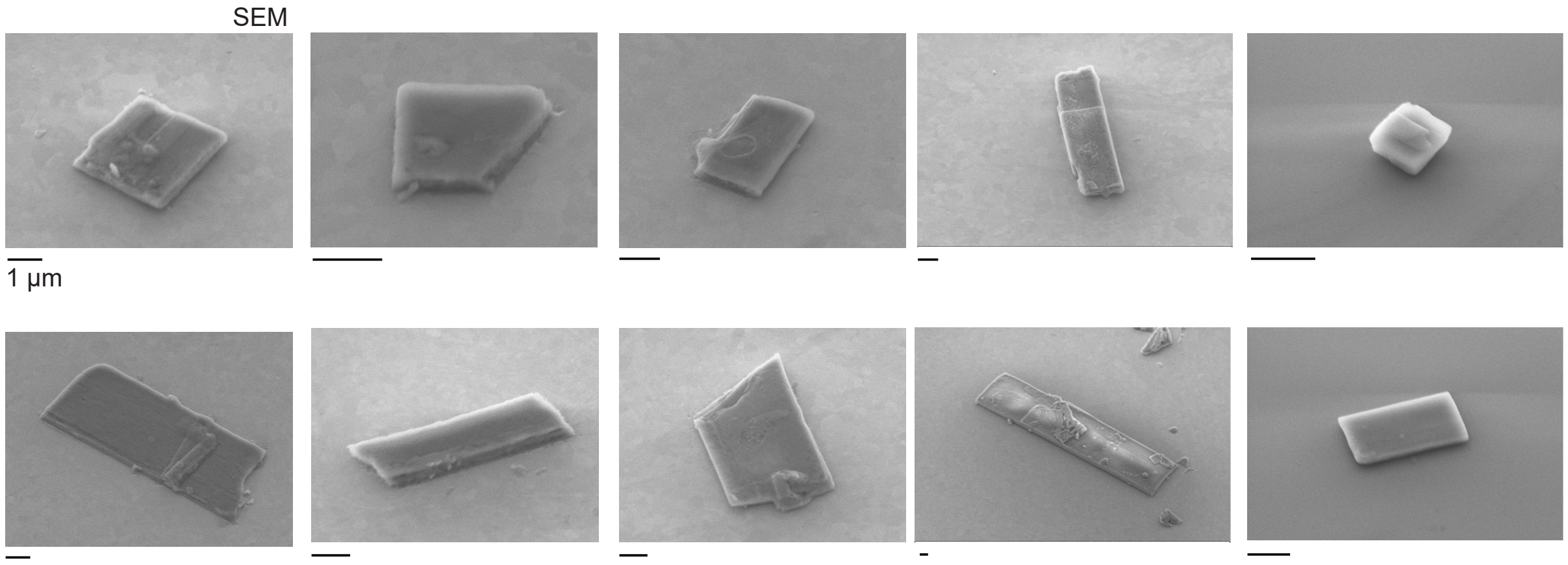

**Supplementary Fig. S9.** SEM images of representative 5IPA3 (n=4) microcrystals, ranging from a few microns to tens of microns in size, on gold substrates. Scale bar, 1 μm.

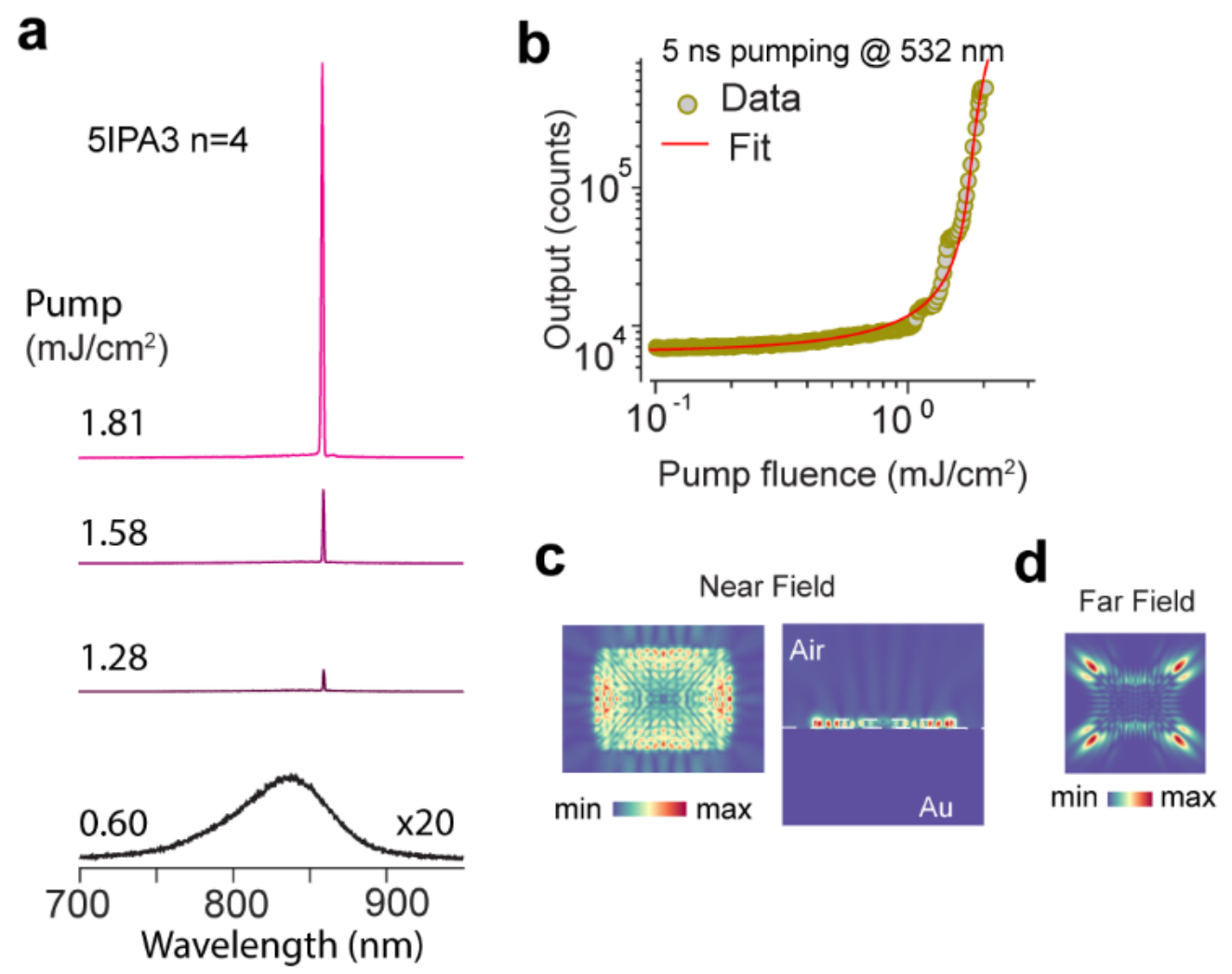

**Supplementary Fig. S10. Representative n=4 plasmonic devices. a,** Output spectra of 5IPA3 n=4 microcrystal at different pump fluences: 0.6, 1.28, 1.58, and 1.81 mJ/cm². **b,** Measured light-in-light-out curve. Dashed line: theoretical fit. **c,** FDTD-simulated electric field amplitude, |E|, of the lasing mode with a cold cavity Q factor of 40. **d,** FDTD simulation for the far-field profile of the lasing mode.

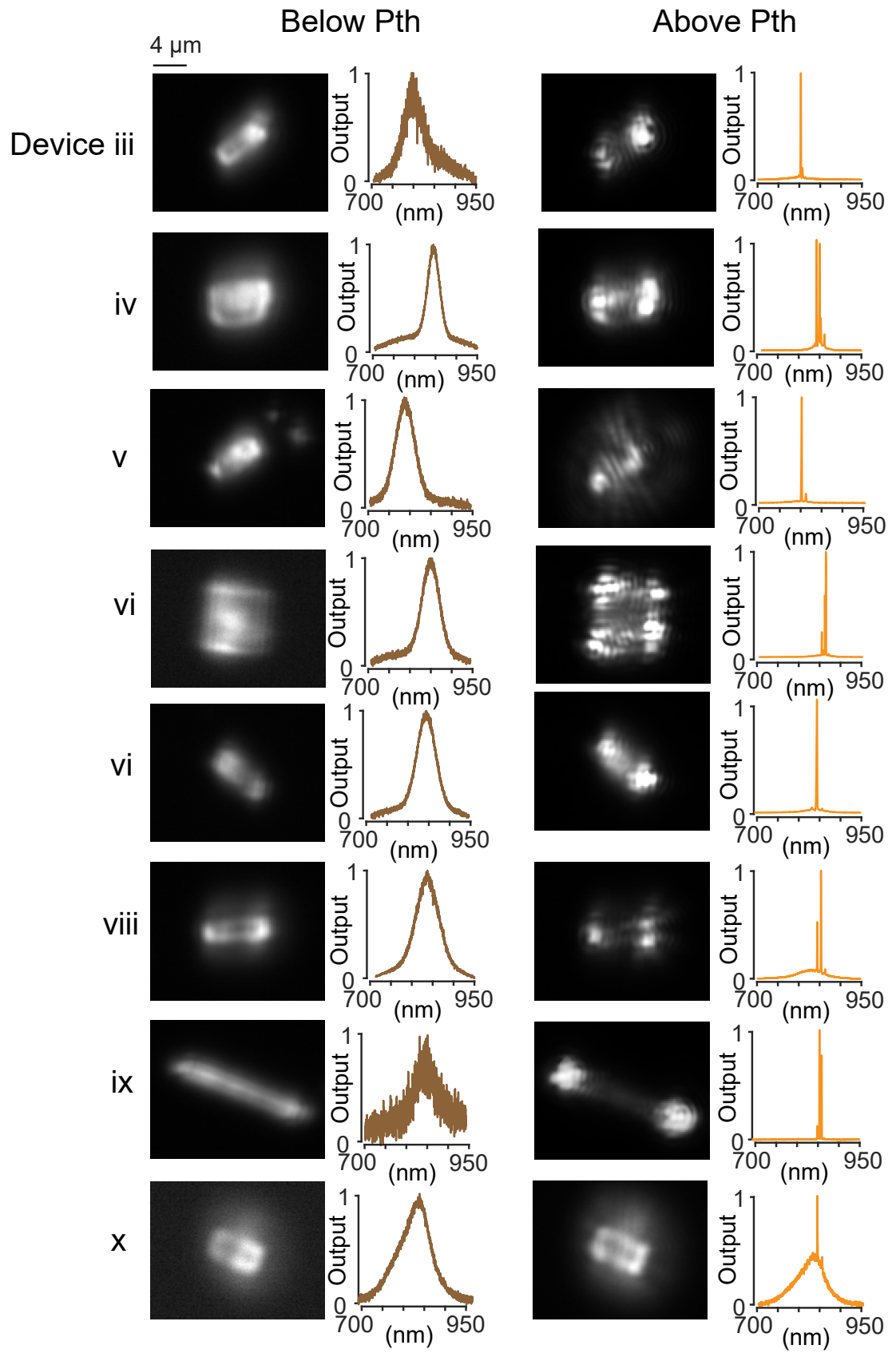

**Supplementary Fig. S11. Shape and spectra of n=3 and n=4 plasmonic laser devices.** Devices of a few micrometers in size. For each device, a representative optical image and emission spectra are shown: (left) at a pump energy of 0.3 (dark brown) times the threshold, and (right) at a pump energy of 1.2 (dark yellow) times the threshold. Scale bar, 4 μm.

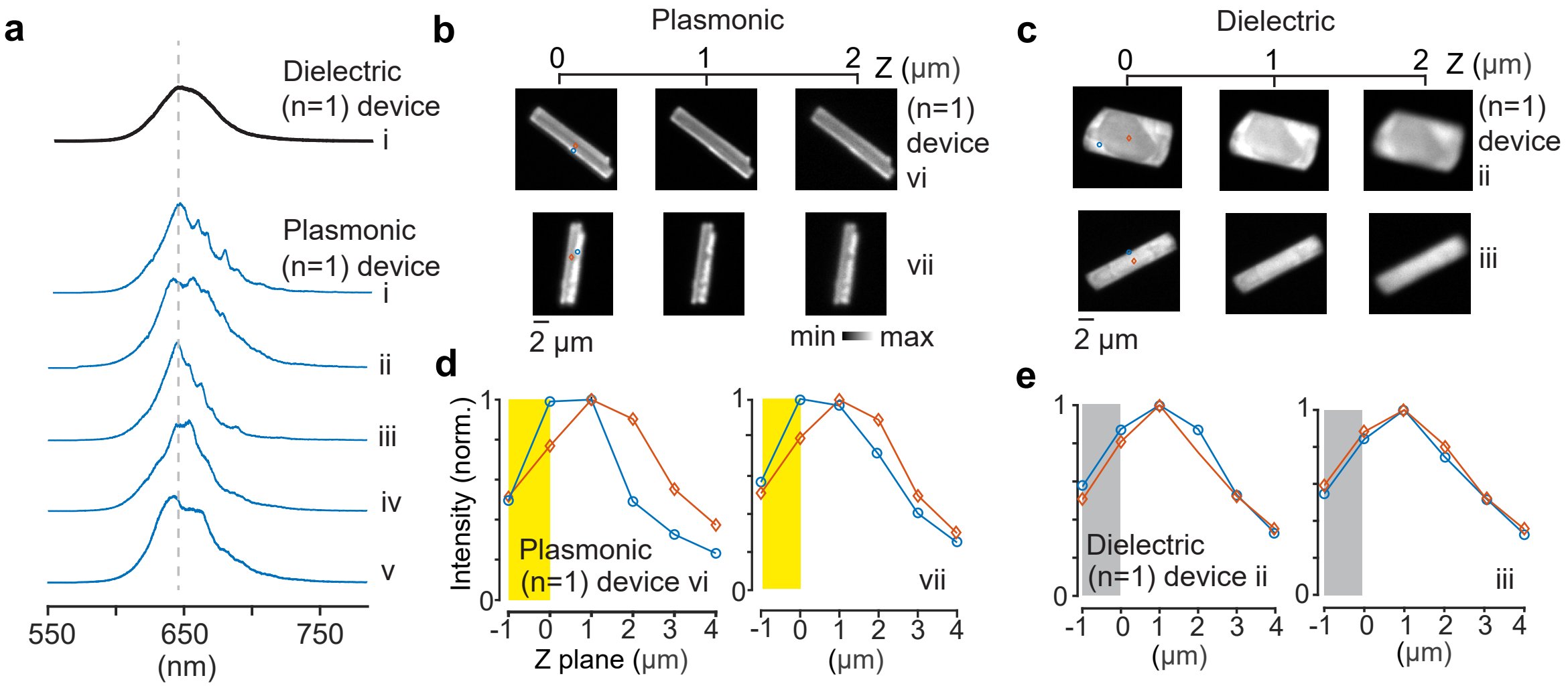

**Supplementary Fig. S12. Purcell-Enhanced Exciton-Surface Plasmon Polariton Coupling in Two-Dimensional n=1 5IPA3 Tin Halide Perovskites Near the Dispersion Asymptote. a,** Fluorescence spectra for the reference dielectric device (black, i) and plasmonic devices (blue, i-v). **(b-c)** Confocal excitation PL images at different focal plane positions (Z) from the substrates for various n=1 microcrystals on different substrates: (b) devices on gold, (c) devices on thermal oxide silicon. Scale bar, 2 μm. **(d-e)** PL intensity measured at bright (blue) and dark (orange) region from the images in (b) and (c).

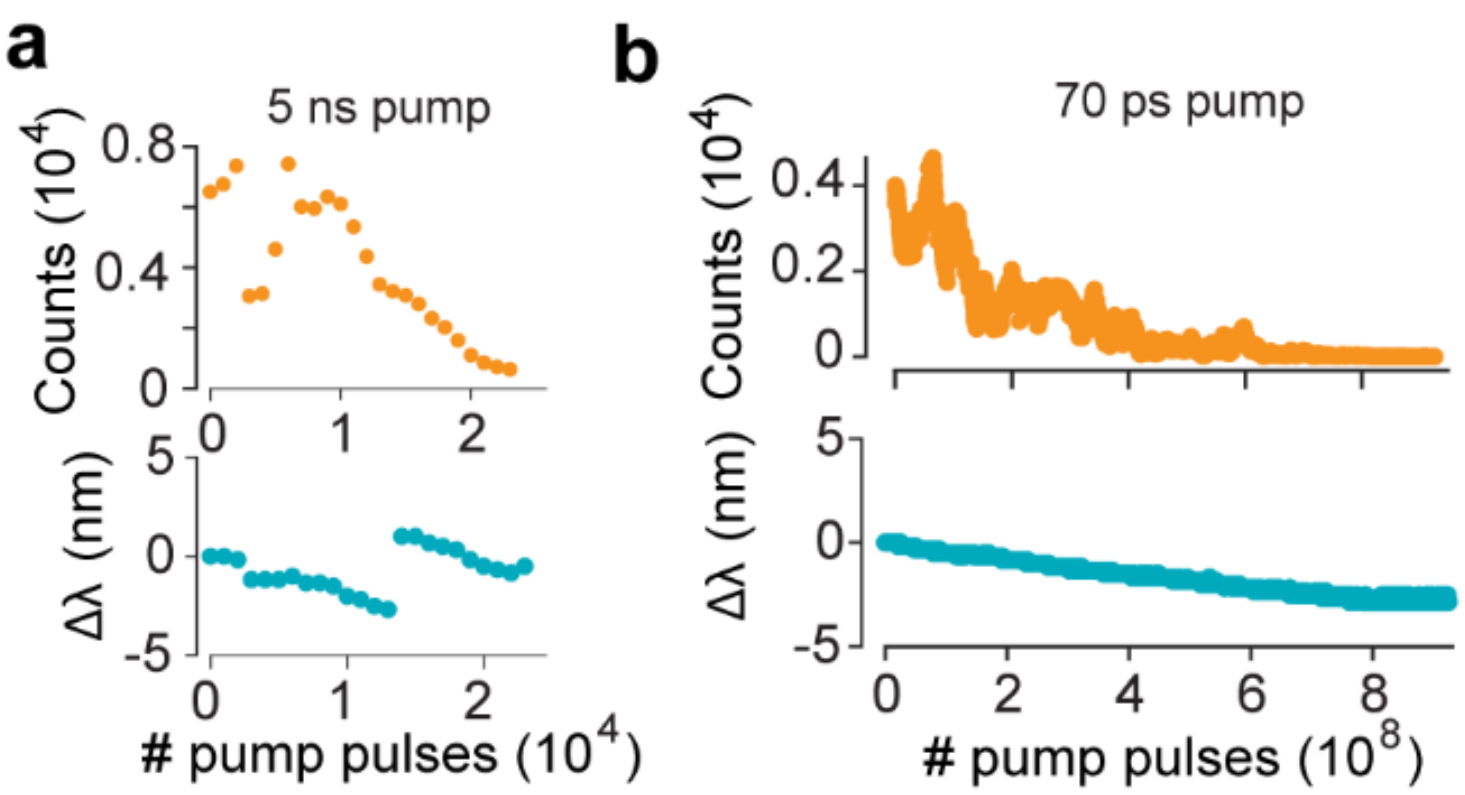

**Supplementary Fig. S13.** The intensity (top) and wavelength (bottom) variations of an n=4 plasmonic device (a) nanosecond pumping and (b) picosecond pumping. Exponential fittings yielded the 1/e parameter for intensity changes as $-1.2 \times 10^{-4}$ shots$^{-1}$ and $-0.9 \times 10^{-4}$ shots$^{-1}$ for the two devices. Linear fittings to wavelength shifts gave slopes of $-3 \times 10^{-4}$ nm/shots and $-2 \times 10^{-4}$ nm/shots.

**Supplementary Table S3. Representative 2D perovskite laser devices to date.**

| | Type (External Cavity) | Materials (Layer number /Wavelength) | Gain-medium size | Pump at threshold (Calculated value) | Pulse Wavelength/Duration (Repetition) | Working Temperature (Surrounding Gas) | Year (Ref) |
|---|---|---|---|---|---|---|---|
| 1 | Microcrystal | $(BA)_2(MA)_{n-1}Pb_nI_{3n+1}$ (n=3 / 630 nm) | Length: ≈11.2 µm Width: ≈11.1 µm Thickness: ≈150 nm | 2.6 µJ/cm$^2$ (32.5 MW/cm$^2$) | 400 nm/80 fs (1 kHz) | 78 K (Vacuum) | 2019 ([1]) |
| 2 | Bulk film (Distributed Feedback) | PEABr based $FAPbBr_3$ (Not reported / 560 nm) | Length: ≈1 mm Thickness: 70−110 nm | 59 W/cm$^2$ | 488 nm/CW | 298 K (Air) | 2020 ([2]) |
| 3 | Bulk film (Distributed Bragg Reflector) | $(PEA)_2(MA, Cs)_{n-1}Pb_nI_{3n+1}$ (n=phase nonpure / 540 nm) | Length: ≈ 1 mm Thickness: 50 nm | 143 µJ/cm$^2$ (4 GW/cm$^2$) | 400 nm/35 fs (1 kHz) | 298 K (Vacuum) | 2021 ([3]) |
| 4 | Microcrystal | $(PEA)_2(MA)_{n-1}Pb_nI_{3n+1}$ (n=3 / 625 nm) | Length: ≈9 µm | 2 GW/cm$^2$ | 800 nm/50 fs (1 kHz) | 173 K (Nitrogen) | 2022 ([4]) |
| 5 | Bulk film (Distributed Bragg Reflector) | $PEA_2PbI_4$ (n=1 / 665 nm) | Length: ≈ 1 mm | 19 µJ/cm$^2$ (23.8 MW/cm$^2$) | 532 nm/800 ps (1 kHz) | 77K (Nitrogen) | 2022 ([5]) |
| 6 | Microcrystal | $(3T^+)_2MASn_2I_7$ (n=2 / 730 nm) | Length: ≈10 µm | 374 µJ/cm$^2$ (2.5 GW/cm$^2$) | 515 nm/150 fs (100 kHz) | 300 K (Vacuum) | 2023 ([6]) |
| 7 | Microcrystal | $(3T^+)_2MASn_2I_7$ (n=2 / 730 nm) | Length: ≈10 µm | 11.2 µJ/cm$^2$ (75 MW/cm$^2$) | 515 nm/150 fs (100 kHz) | 83 K (Vacuum) | 2023 ([6]) |
| 8 | Microcrystal | $(3T^+)_2MAPb_2I_7$ (n=3 / 620 nm) | Length: ≈20 µm | 30 µJ/cm$^2$ (150 MW/cm$^2$) | 400 nm/200 fs (1 kHz) | 150 K (Vacuum) | 2023 ([7]) |
| 9 | Microcrystal | $(BrCA3)_2PbBr_4$ (n=1 / 630 nm) | Length: ≈100 µm | 17 µJ/cm$^2$ (2.5 GW/cm$^2$) | 400 nm/100 fs (1 kHz) | 88K (Argon) | 2024 ([8]) |
| 10 | Bulk film (Lattice plasmon cavity) | $(PEA)_2Cs_{n-1}Pb_nBr_{3n+1}$ (n=5 / 518 nm) | Length: ≈1.5 mm | 11.2 mJ/cm$^2$ (112 GW/cm$^2$) | 400 nm/100 fs (1 kHz) | 298K (Not reported) | 2025 ([9]) |
| 11 | Microcrystal on thermal oxide silicon | $(5IPA_3)_2(MA)_3Sn_4I_{13}$ (n=4 / 850 nm) | Length: ≈ 4 µm | 1000 µJ/cm$^2$ (200 kW/cm$^2$) | 532 nm/5 ns (10 kHz) | 298 K (Room Air) | 2025 (This work) |
| 12 | Microcrystal on gold | $(5IPA_3)_2(MA)_3Sn_3I_{10}$ (n=3 / 800 nm) | Length: ≈4 µm | 750 µJ/cm$^2$ (150 kW/cm$^2$) | 532 nm/5 ns (10 kHz) | 298 K (Room Air) | 2025 (This work) |

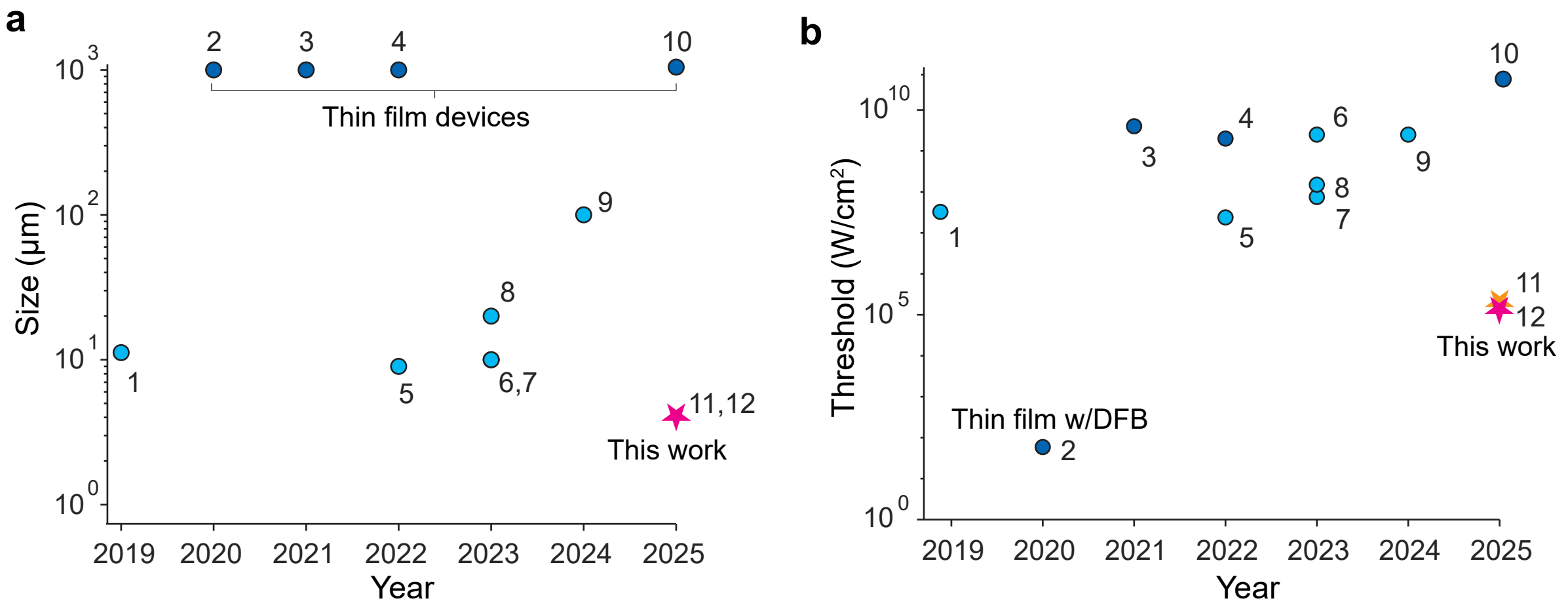

**Supplementary Fig. S14. (a-b) a, Device volume and b, threshold of various 2D perovskite-based lasers.** Each data point is labeled with the device number used in Supplementary Table S3.